\documentstyle[aps,epsf]{revtex}

\begin{document}
\title{Modified Newtonian Dynamics In Dimensionless Form}
\author{
W. F. Kao\thanks{%
gore@mail.nctu.edu.tw} \\
Institute of Physics, Chiao Tung University, Hsinchu, Taiwan}
\maketitle

\begin{abstract}
Modified Newtonian dynamics proposed  that gravitational field
needs modifications when the field strength $g$ is weaker than a
critical value $g_0$. This has been shown to be a good candidate
as an alternative to cosmic dark matter. There is another way to
look at this theory as a length scale dependent theory. One will
show that modification of the Newtonian field strength depends on
the mass distribution and the coordinate scale of the system. It
is useful to separate the effective gravitation field $g(r)$ into
a small scale (or short-distance )$g_s$ field and a large scale
(or a long-distance) $g_l$ field that should be helpful for a
better understanding of the underlying physics. The effective
potential is also derived.
\end{abstract}



\section{INTRODUCTION}
Modified Newtonian dynamics (MOND) was proposed by Milgrom (1983)
that gravitational field needs modifications when the field
strength $g$ is weaker than a critical value $g_0$. This has been
shown to be a good candidate as an alternative to cosmic dark
matter (Sanders 2001). The phenomenological foundations for MOND
are based on two observational facts: (1) flat asymptotic rotation
curve, (2) the successful Tully-Fisher (TF) law, $M \sim V^\alpha$
(Tully \& Fisher 1977) for the relation between rotation velocity
and luminosity in spiral galaxies. Here $\alpha$ is close to 4.

In this paper, one proposes a different view of MOND by looking at
the physics related by the mass distribution and the coordinate
scale of the system. In addition, one finds it useful to separate
the effective gravitation field $g(r)$ into a small scale (or
short-distance )$g_s$ field and a large scale (or a long-distance)
$g_l$ field that should be helpful for a better understanding of
the underlying physics. The effective potential is obtained.
Possible relation with the induced gravity model is also
discussed.

If dark matter does not exit, rotation curve observations indicate
that $g$ behaves as $1/r$, a typical $2$-dimensional attraction
field, at large  distance greater than the galactic scale around
$10^5$ lightyears. At short distance, $g$ goes like $1/r^2$, a
typical behavior of a $3$-dimensional attraction field. Therefore,
it is also very interesting to study the changing pattern of the
\emph{effective dimension} $d(r)$ which will be defined as a
function of the physical scale $r$. One will plot this $d(r)$
functions hoping for a better understanding of the changing
pattern of the \emph{effective dimension}. Possible speculation
with the Kaluza-Klein theory will be discussed in this paper too.

We will start by showing how to simplify the expression of MOND
field strength by writing it with suitable scale parameters. Our
aim is to write all physical quantities in dimensionless form with
suitable physical units.

First of all, it was pointed out (Milgrom 1983) that there exists
a critical acceleration parameter $g_0 = 1.2 \times 10^{-8} {\rm
cm s}^{-2}$ characterizing the turning point of the effective
power law associated with the gravitational field in MOND.
Gravitational field of the following form was suggested
\begin{equation} \label{gm}
g \cdot \mu ({g \over g_0}) =g_N
\end{equation}
with a function $\mu$ considered as a modified inertial. Here
$g_N$ is the Newtonian gravitational field produced by certain
mass distribution. Milgrom argues that
\begin{equation}
\mu (x) = {x  \over  \sqrt{1+x^2}}
\end{equation}
provides a best fit with many existing observation data including
the rotational curve of many spiral galaxies. Observations are
selected mostly from the most reliable $21$ cm hydrogen line
measurements. The 21 cm hydrogen line is the photon emission due
to the superfine energy difference resulting from the magnetic
dipole-dipole interactions between the cored proton and orbiting
electron. The existing neutron hydrogen all around the invisible
region of spiral galaxies provides better estimates on the
rotation curves beyond the visible region of spiral galaxies.

Milgrom also points out that the critical parameter $g_0$ is close
to the value of $cH_0 \sim 5 \times 10^{-8} cm s^{-2}$. It is also
the gravitational field produced by an $m=200 MeV$ massive
particle at its Compton surface. Indeed, one finds that $g_{200
MeV}=Gm/\lambda=Gm^3c^2/h^2 \sim 1.2 \times 10^{-8} cm s^{-2}$.
Milgrom hence speculates that MOND could have to do with the
quantum theory of the universe.

Milgrom's original idea is that the gravitational field needs
modification when $g$ is weaker than critical field strength
$g_0$. It is interesting to find that $r_0 \sim 49700 ly$ if we
define the scale length $r_0$ via the following equation
\begin{equation} \label{g0}
g_0 \equiv {G M_0 \over r_0^2}
\end{equation}
with $M_0 \equiv 2 \times 10^{11} M_s$ and $M_s$ the solar mass.
Note that one takes $M_0$ close to the order of magnitude of a
typical galaxy similar to our Milky Way. Taking a different value
of $M_0$ will only change the value of $r_0$. Selecting different
value of $M_0$ and hence $r_0$ will not affect our arguments in
this paper. One chooses $M_0$ and hence the corresponding $r_0$
simply to set them as units of scale. Physics will not be affected
by the choice of unit.

One observes, however, that $r_0$ is pretty close to the radial
size of the visible boundary of our Milky Way if $M_0$ is chosen
close to the total visible mass of Milky Way. This is certainly
true as modification is needed only at large scale according to
MOND. Starting next section, one will try to write the modified
field strength $g$ in a dimensionless form with the help of the
parameters $M_0$, $r_0$ and $g_0$. The idea is to write $M$, $r$
and $g$ in units of $M_0$, $r_0$ and $g_0$ respectively.

\section{Modified Newtonian Dynamics}

Milgrom argues that one can take the effective inertial $\mu$ as
$\mu (x) = x / \sqrt{1+x^2} $. Therefore, one obtains
\begin{equation} \label{g}
{(g/g_0)^2 \over \sqrt{1+(g/g_0)^2} }= g_N/g_0
\end{equation}
by dividing both sides of the equation [\ref{gm}] by the critical
parameter $g_0$. One can therefore write above equation as
\begin{equation}
{ g^2 \over \sqrt{1+g^2} }= g_N
\end{equation}
with $g$ and $g_N$ written in unit of $g_0$. This makes $g$ and
$g_N$ dimensionless from now on.

Throughout this paper, we will focuss on the study of the system
with a Newtonian attraction of the form $g_N =Gm/r^2$. This is the
field strength at a radial distance $r$ from a spherical
distributed matter with total mass $m$. System with different mass
distribution can be obtained straightforwardly. Note that ${Gm /
(r^2g_0)} = {mr_0^2 / (M_0 r^2)}$. Therefore, one can write $g_N
=m/r^2$ if $g_N$, $r$ and $m$ are written in unit of $g_0$, $r_0$
and $M_0$ respectively. Therefore, the modified field strength $g$
can be written as
\begin{equation}
{ g^2 \over \sqrt{1+g^2} }=  {m \over r^2} \label{01}
\end{equation}
One can further suppress the parameter $m$ by defining $r_c \equiv
r_0 \sqrt{m}$ and write $r$ as a dimensionless coordinate variable
in unit of $r_c$ instead of $r_0$. As a result, one can write
above equation in a very compact form
\begin{equation}
{ g^2 \over \sqrt{1+g^2} }=  {1 \over r^2} \label{1}
\end{equation}
in terms of above dimensionless parameters. To emphasize again,
one writes
\begin{eqnarray} \label{dim}
{m \over M_0} &\to& m, \nonumber \\
{g \over g_0} &\to& g , \nonumber \\
{r \over r_c}={ r \over \sqrt{m}r_0} &\to& r
\end{eqnarray}
in order to simplify the expression of the modified equation and
make $g$, $g_N$, and $r$ dimensionless for convenience. It is
straightforward to restore all dimension parameters when one needs
to evaluate any corresponding physical values.

\begin{figure}[h]
\epsfxsize=10cm \centerline{\epsffile{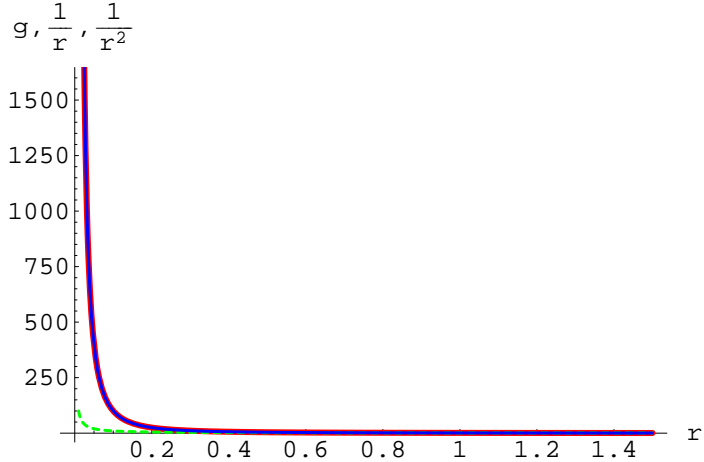}} \caption{$g(r),
1/r$ and $1/r^2$ plotted as thick solid line, dashed line, and
thin solid line respectively. Note that $g$ and $r$ are
dimensionless quantities defined in Eq. [\ref{dim}]. }
\label{FIG1}
\end{figure}

\begin{figure}[h]
\epsfxsize=10cm \centerline{\epsffile{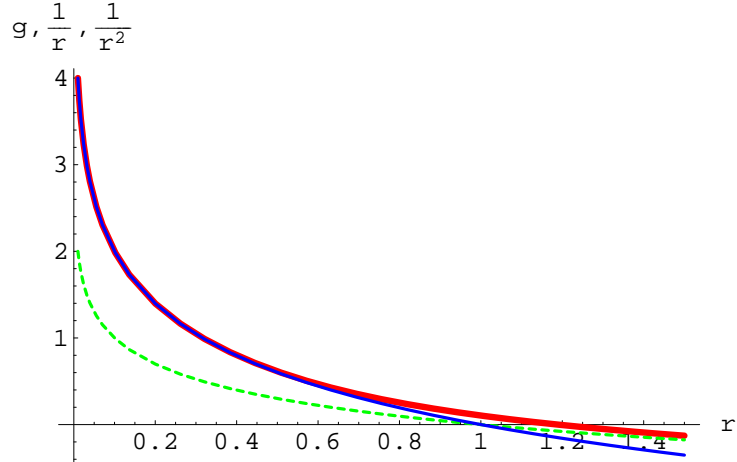}} \caption{Logarithm
 of $g(r), 1/r$ and $1/r^2$ plotted as thick solid line, dashed line,
and thin solid line respectively. Note that $g$ and $r$ are
dimensionless quantities defined in Eq. [\ref{dim}]. }
\label{FIG2}
\end{figure}

First of all, Eq. (\ref{1}) can be solved to write $g(r)$ as a
function of $r$:
\begin{equation} \label{2}
g= {\sqrt{ 1+ \sqrt{ 1+4r^4} } \over  \sqrt{2} r^2}={1 \over  r}
\exp [ { \sinh ^{-1} (2r^2)^{-1} \over 2}] .
\end{equation}
One can plot the functions $g(r)$ and compare it with the
functions $1/r$ and $1/r^2$ for a better picture of its small
scale and large scale behaviors. Indeed, Figure \ref{FIG1} shows
the behavior of $g(r), 1/r$, and $1/r^2$ together for comparison.
Because the difference at short distance scale is too small for
comparison, one also plot it in logarithmic scale. Figure
\ref{FIG2} was plotted as logarithm of these functions to enlarge
the small difference among these functions in the short distance
region.

It is apparent that $g(r)$ goes like $1/r^2$ at short distance
scale where $r \ll 1$. On the other hand, $g(r)$ goes like $1/r$
at large distance scale where $r \gg 1$. One can also easily tell
from the figures that they agree with our analytic approximation
at both the regions $r \ll 1$ and $r \gg 1$ one obtained earlier.

More importantly, writing $g(r)$ as a function of $r \to
r/(\sqrt{m}r_0)$ clearly shows that the Newtonian gravitational
field receives significant modification only when $r >1$, or
equivalently, $r >r_c \equiv \sqrt{m}r_0$. This is in fact
equivalent to the assumption made by Milgrom that significant
modification is required when $g < g_0$. The picture looking at
the physics related to the physical scale is however easier for a
better understanding of possible underlying physics.

Indeed, one easily finds that, if Eq. (\ref{2}) is the correct
gravitational field for all scales, the effective dimension of the
system seems to decrease from 3 to 2 as the scale $r$ increases.
Geometric dimension seem to decrease if one increase the physical
scale of interest. The major difference is that the critical scale
depends not only on $r_0=\sqrt{GM_0/g_0}$ but also proportional to
$\sqrt{m}$, the square-root of the total mass of the system. One
recalls that Kaluza-Klein theory proposes that observable physical
dimension of a system increases as the physical length scale
decreases, or equivalently, the energy scale increases. One will
be able to observe higher dimensional effect at shorter distance
scale according to Kaluza-Klein's original idea.

The situation here is quite similar. When the physical length
scale increases, the effective physical dimension appears to
decreases from 3 to 2. Moreover, the critical length scale $r_c
\equiv \sqrt{m}r_0$ depends on the total mass of the system. If
$m$ increases, equivalent to the increasing of total energy of the
system, MOND modification is only apparent at larger scale.
Imagine one approaches a system from a large distance area. One
will be able to see the effect of one more dimension as one
approaches closer to the center of the system. Increasing $m$ will
make one easier to see the differences at the same physical length
scale. This agrees with the central idea of Kaluza-Klein approach.
One can possibly say that energy or mass opens up more dimensions
if Eq. (\ref{2}), or any similar expression of the MOND field
strength, is the correct equation for the gravitational field.

Note that one can also tell from Eq. \ref{2} that one can expand
the effective theory in $r$ if one needs analytic form of the
theory either at short or large scale. The expansion in $r$ is in
fact the expansion in $r/r_c$. When the scale length $r/r_c$ is
small, it returns to the conventional Newtonian model. It will
receive significant modification when $r$ is large compared to
$r_c$. Therefore, attention should be addressed to the physical
origin of the parameter $r_c$.

In order to take a more close look at the changing pattern of the
above $g$, one can split it into two different parts:
\begin{equation}
g=g_l+g_s= { 2\sqrt{2} r^2  \over  \sqrt{1+4r^4} \sqrt{ 1+ \sqrt{
1+4r^4} } } + { \sqrt{ 1+ \sqrt{ 1+4r^4} }  \over  \sqrt{2} r^2
 \sqrt{1+4r^4} }.
\end{equation}
Note that the first term is $g_l$ that goes like $1/r$ when $r \gg
1$, while the second term is $g_s$ that goes like $1/r^2$ when $r
\ll 1$. Therefore it is easy to see that $g_l$ and $g_s$
represents the long distance and short distance field strength of
the $g$ respectively. One is hoping that successful separation of
$g$ May shed more light in the search of the underlying theory.

In fact, one can show that
\begin{eqnarray}
g_l &\to& 2r^2 \\
g_s &\to& {1 \over r^2} - {3 \over 2} r^2
\end{eqnarray}
and hence,
\begin{equation}
g \to  {1 \over r^2} + {1 \over 2} r^2 = {1 \over r^2} (1 + {1
\over 2} r^4 )
\end{equation}
as $r \to 0$. This indicates that the first order correction to
the short-range gravitational field $g$ is $O(r^2)$ in comparison
to the $1/r^2$ Newtonian interaction. Hence the first order
correction is fourth order in $r$. This correction is very tiny if
one takes the mass $m$ to be the solar mass $M_s$. In this case
the critical radius $r_c ~ 0.111 ly$. Hence, the first order
correction $\delta g$ at a distance from our sun near $r=10$ AU,
would be around a small difference near $4.16 \times 10^{-12}
\times g_{10}$. Here $g_{10}$ denotes the Newtonian gravitational
field at a distance $r=10$ AU from our sun. This would probably
large enough to be measured in the near future.

\begin{figure}[h]
\epsfxsize=10cm \centerline{\epsffile{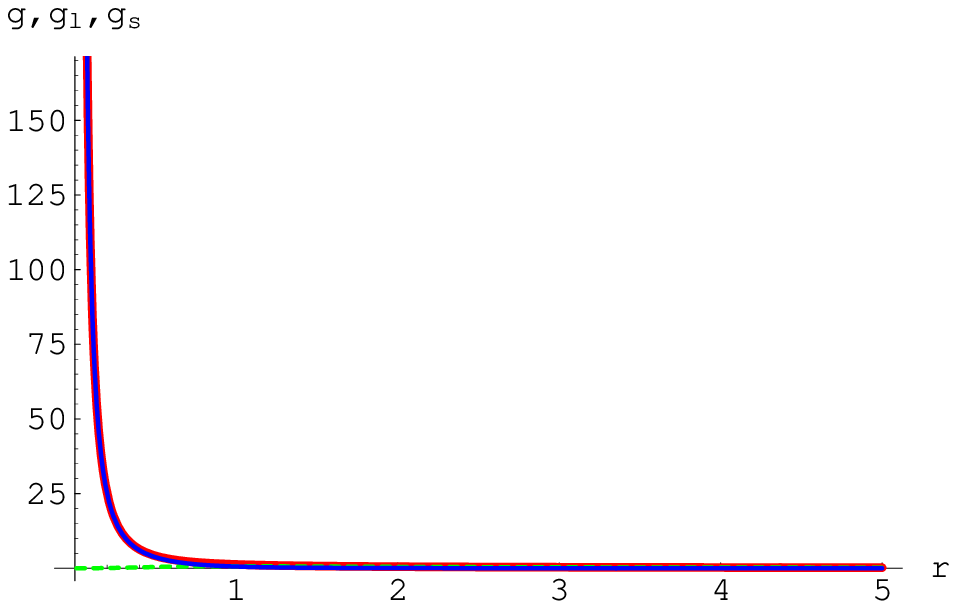}} \caption{ $g(r),
g_l(r)$ and $g_s(r)$ plotted as thick solid line, dashed line, and
thin solid line respectively. Note that $g$ and $r$ are
dimensionless quantities defined in Eq. [\ref{dim}]. }
\label{FIG3}
\end{figure}

\begin{figure}[h]
\epsfxsize=10cm \centerline{\epsffile{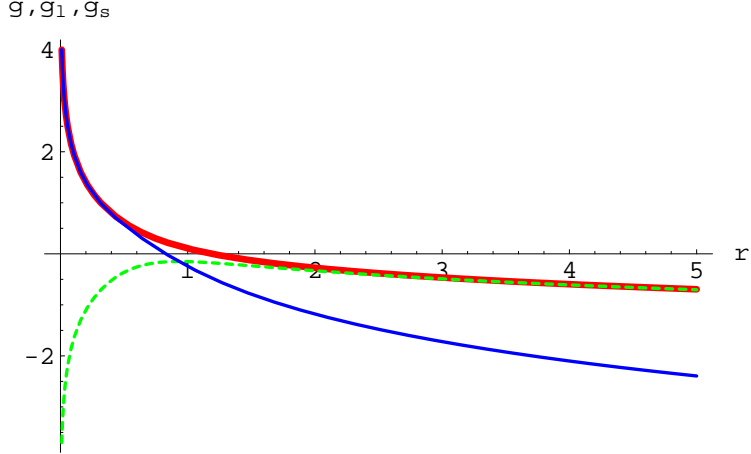}} \caption{Logarithm
of  $g(r), g_l(r)$ and $g_s(r)$ plotted as thicker solid line,
dashed line, and solid thin line respectively. Note that $g$ and
$r$ are dimensionless quantities defined in Eq. [\ref{dim}]. }
\label{FIG4}
\end{figure}

In Figure \ref{FIG3} one plots $g(r), g_l(r)$, and $g_2(r)$ for
comparison. We also plot the logarithm of these functions in Fig.
\ref{FIG4} in order to signify the small difference in short
distance.

In addition, one can also compute the changing rate of the
gravitational field for the following equation of running
gravitational field
\begin{equation} \label{rg}
\partial_r g = -2 g^2{ (g^2+1)^{3/4} \over g^2+2}.
\end{equation}
If we define $k=1/r$ as a parameter simulating the effect of the
conjugated momentum, one can also show that the changing rate
$\partial_k g(k)$ obeys the following equation
\begin{equation}  \label{rgk}
\partial_k g =2 {(g^2+1)^{5/4} \over g^2+2}
\end{equation}
These equations play the role similar to the renormalization group
equation for a quantum field theory. In addition, Eq. [\ref{rg}]
also states that $g(r)$ decreases monotonically as $r$ increases.
Similarly, Eq. [\ref{rgk}] indicates that $g(k)$ increases
monotonically as $k$ increases.

\section{Effective Potential and effective dimension}

One can integrate $g$ for the effective potential.  After some
algebra, One can show that the effective potential $V_l \equiv -
\int_\infty^r {\bf g}_l \cdot d{\bf r} = \int^r g_l dr$ and
similarly for $V_s \equiv \int^r g_s dr$ can be evaluated directly
to give
\begin{equation}
V_l=  \ln { \sqrt{1+ \sqrt{1+4r^4}} \over \sqrt{2}}-
\sum_{n=1}^{\infty} { \pi_{k=1}^n (4k-3)  \over 2n \cdot n! 2^n
(1+\sqrt{ 1+4r^4} )^n } ,
\end{equation}
and
\begin{equation}
V_s =  - { \sqrt{ 1+ \sqrt{ 1+4r^4} } \over\sqrt{2} r }
\end{equation}
One can verify directly that $V_l'=g_l$ and $V_s'=g_s$ and prove
that above equations are indeed correct up to an irrelevant
integration constant. In fact, it is difficult to specify this
constant of integration in the conventional approach which take
$V(r \to \infty) \to 0$. This is because the effective potential
in fact diverges at spatial infinity due to the logarithm behavior
of the dominating 2D-like potential.

\begin{figure}[h] \epsfxsize=10cm
\centerline{\epsffile{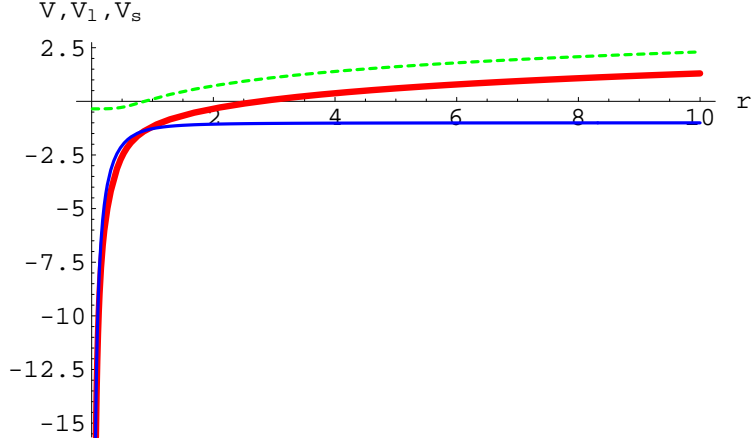}} \caption{ $V(r), V_l(r)$ and
$V_s(r)$ plotted as thick solid line, dashed line, and thin solid
line respectively.  } \label{FIG5}
\end{figure}

In order to take a look at the form of these effective potential ,
one plots $V, V_l,$ and $V_s$ together in Fig. \ref{FIG5}. It is
easy to find that $V_l$ has a flat plateau area when $r \to 0$
indicating that $g_l$ contributes little to $g$ at short distance.
Similarly, one finds that the slope of $V_s$ becomes constant as
$r  \gg 1$  indicating that $g_s$ is significant only at short
distance scale. Note that, the series sum in $V_s$ converges
quickly. One takes $n=100$ when we plot Fig. \ref{FIG5}.

One knows that the four dimensional space time is modified when
the physical scale becomes small in the Kaluza-Klein theory.
Recent evidences in MOND research seems to imply that the
effective dimension of our universe decreases as the length scale
increases. It is possible that Kaluza-Klein approach could play an
important, but quite different, role for large scale system. The
observed rotation curve of spiral galaxies indicates that the
gravitational attraction goes as $1/r$ in all directions without
the existence of any cosmic dark matter. This is, however, quite
different from the conventional geometry of any two dimensional
space. But the success of MOND seems to indicate a new direction
of dimensional effect.

It is then very interesting to study how the effective dimension
decreases as the observation scale increases. Therefore, one
manages to define an effective dimension in this paper hoping to
find our way to the discovery of the underlying theory.

\begin{figure}[h]
\epsfxsize=10cm \centerline{\epsffile{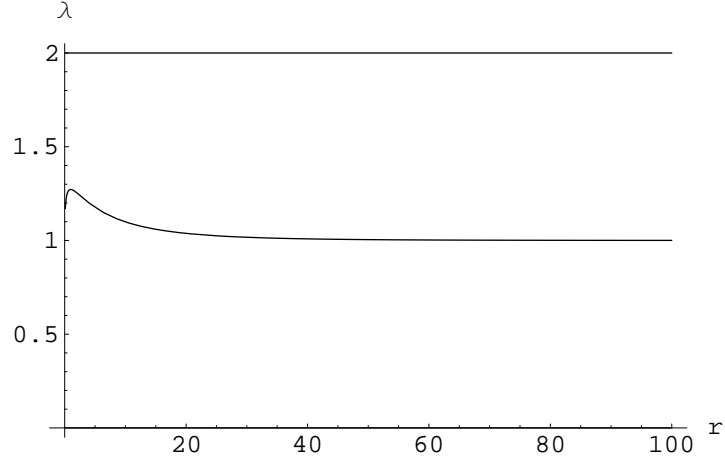}} \caption{
$\Lambda$-shaped smooth function one needs is plotted as function
$r$. Note that $g$ and $r$ are dimensionless quantities. }
\label{FIG41}
\end{figure}

\begin{figure}[h]
\epsfxsize=10cm \centerline{\epsffile{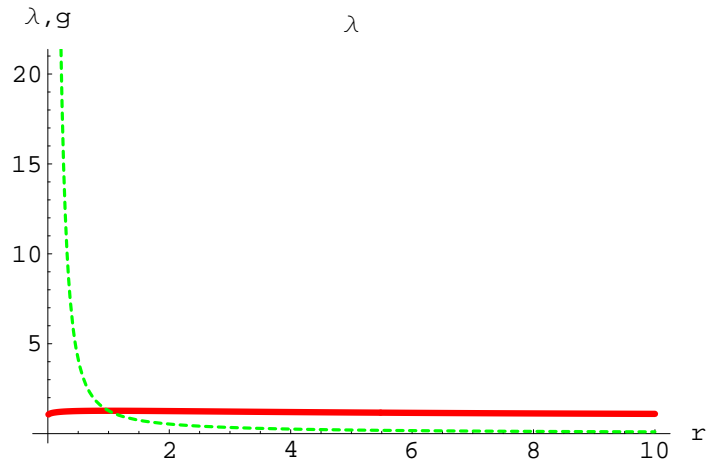}}
\caption{$\lambda$ and $g(r)$ are plotted as thick solid line,
dashed line respectively. } \label{FIG9}
\end{figure}

Indeed, one can define the effective dimension $d(r)$ by the
definition
\begin{equation} \label{dg}
g = {1 \over r^{d-1}}
\end{equation}
such that $d \to 3$ as $r \to 0$ and $d \to 2$ as $r \to \infty$.
This is a naive way to define an effective dimension such that $d$
will signify the geometric dimension as $3$-dimension and
$2$-dimension respectively in a close way. It is then
straightforward to show that
\begin{equation}
r^{d-3} = { \sqrt{ 1+ \sqrt{ 1+4r^4} } \over\sqrt{2} }
\end{equation}
and hence,
\begin{equation}
d = 2 - { \sinh ^{-1} {1 \over 2r^2} \over 2 \ln r} .
\end{equation}
Note, however, that this definition of $d(r)$ fails to tell us
anything when $r \to 1$. The reason is that $d$ is ill-defined at
$r=1$ which is an obvious result of Eq. \ref{dg}. One resolution
to this problem is to introduce a $\Lambda$-shaped smooth function
$\lambda(r)$ that peaks at $r=1$ reaching the value $\sqrt{ 1+
\sqrt{ 5} } / \sqrt{2} \sim 1.27$ and reaching $1$ both at $r=0$
and $r \to \infty$ such that
\begin{equation}
\lambda(r) r^{d-3} = { \sqrt{ 1+ \sqrt{ 1+4r^4} } \over\sqrt{2} }
\end{equation}
renders $d(r)$ well defined at $r=1$. Indeed, one can show that
\begin{equation}
d(r) = 3-{
 \ln { \sqrt{ 1+ \sqrt{ 1+4r^4} } \over \sqrt{2} } -\ln \lambda(r)
\over \ln r
 }.
\end{equation}
In short, the prescribed $\Lambda$-shaped function $\lambda(r)$
absorbs the discrepancy of the function $d(r)$ at $r=1$ in order
to make the $d$ function well-defined everywhere including $r=1$.
For simplicity, one will choose
\begin{equation}
 \lambda(r)= 1+ [{\sqrt{
1+ \sqrt{ 5} } \over  \sqrt{2}} -1]\sqrt{r} \exp [1-\sqrt{r} ].
\end{equation}
One can easily show that this function peaks at $r=1$ reaching the
value $\sqrt{ 1+ \sqrt{ 5} } / \sqrt{2}$ and reaches $1$ both at
$r=0$ and $r \to \infty$ as required. And it has little effect
elsewhere. This is exactly the function shown in Fig. \ref{FIG41}.
In addition, Figure \ref{FIG9} shows $\lambda(r)$ and $g(r)$
together for comparison.

\begin{figure}[h]
\epsfxsize=10cm \centerline{\epsffile{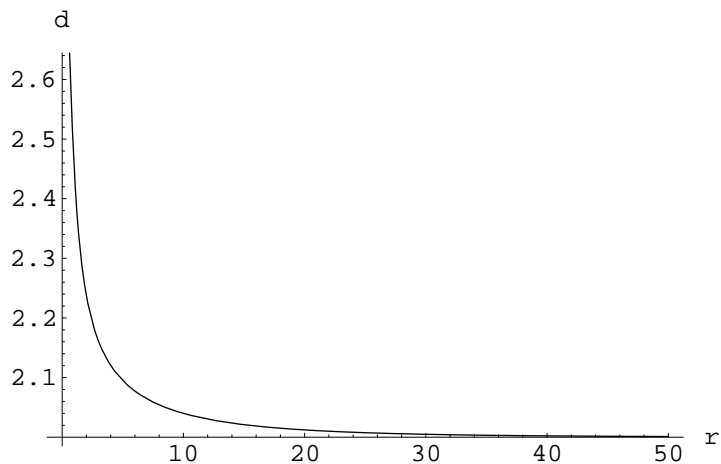}} \caption{ effective
dimension $d(r)$ is plotted as function $r$. } \label{FIG6}
\end{figure}

Therefore, one can plot the function of effective dimension $d(r)$
in Fig. \ref{FIG6}

Note that the effective dimension function $d(r)$ does not
represent the physical dimension of the theory. It only tells us
roughly and effectively how the gravitational field $g$ varies as
$r$ increases. In one aspect, one could view the potential as
distortion of the geometry and the effective potential as the
stiffness of the distorted geometry. Since the $1/r$ effect holds
in all direction at large scale, not just along the disk plane,
this behavior is quite different from simple geometric and
dimensional deformation. All evidences indicate this sort of
geometry deserves more attention.

\section{conclusion}
In this paper, one proposes a different view of MOND by looking at
the physics related by the mass distribution and the coordinate
scale of the system. In addition, one finds it useful to separate
the effective gravitation field $g(r)$ into a small scale (or
short-distance )$g_s$ field and a large scale (or a long-distance)
$g_l$ field that should be helpful for a better understanding of
the underlying physics. The effective potential is obtained
accordingly.

We have also studied the changing pattern of the \emph{effective
dimension} $d(r)$ which will be defined as a function of the
physical scale $r$. One plots this $d(r)$ functions hoping for a
better understanding of the changing pattern of the
\emph{effective dimension}. Possible speculation with the
Kaluza-Klein theory is also discussed in this paper too.

{\bf \large Acknowledgments} This work is supported in part by the
National Science Council of Taiwan.

\vspace{ 2cm}


{\bf \large References}

\begin{thebibliography}{}

\bibitem []{jdb88} Bekenstein, J.D. 1988, in
    {\it Second Canadian Conf. on General Relativity
    and Relativistic Astrophysics}, eds. Coley, A., Dyer, C., Tupper, T.,
    World Scientific, Singapore, p.487
\bibitem []{bm84} Bekenstein, J.D.
    \& Milgrom, M. 1984, ApJ, 286, 7 (BM)
\bibitem []{bs94} Bekenstein, J.D. \&
    Sanders, R.H. 1994, ApJ, 286, 7
\bibitem []{br92} Broeils, A.H. 1992, PhD Dissertation,
    Univ. of Groningen
\bibitem []{cvg91} Casertano, S.
    \& van Gorkom, J.H. 1991, AJ, 101, 1231
\bibitem []{def01} Deffayet, C. 2001, Phys.Lett. B 502, 199
\bibitem [] {fj} Faber, S.M., \& Jackson, R.E. 1976,
    ApJ, 204, 668
\bibitem []{faleal98} Falco, E.E., Kochanek, C.S.,
    Munoz, J.A. 1998, ApJ, 494, 47
\bibitem []{fel84} Felten, J.E. 1984, ApJ, 286, 38
\bibitem [] {fish} Fish, R.A. 1964, ApJ, 139, 284
\bibitem []{free74} Freeman, K.C. 1970, ApJ, 160, 811
\bibitem [] {heal00} Hanany, S. et al. 2000,
    ApJ, 545, L5
\bibitem []{leal00} Lange, A.E. et al. 2001,
     Phys.RevD, 63, 042001
\bibitem [] {mcdb98a} McGaugh, S.S., de Blok,
    W.J.G. 1998a, ApJ, 499, 66
\bibitem [] {mcdb98b} McGaugh, S.S., de Blok,
    W.J.G. 1998b, ApJ, 508, 132
\bibitem []{mcg99} McGaugh, S.S. 1999, ApJ, 523, L99
\bibitem []{mcg00} McGaugh, S.S. 2000, ApJ, 541, L33
\bibitem []{mgeal00}
    McGaugh, S. S., Schombert, J. M., Bothun, G. D.,
    de Blok, W. J. G. 2000, ApJ, 533, 99
\bibitem []{m83a} Milgrom, M. 1983a, ApJ, 270, 365
\bibitem []{m83b} Milgrom, M. 1983b, ApJ, 270, 371
\bibitem []{m83c} Milgrom, M. 1983c, ApJ, 270, 384
\bibitem []{ml84} Milgrom, M. 1984 ApJ, 287, 571
\bibitem []{m94} Milgrom, M. 1994, Ann.Phys, 229, 384
\bibitem []{m99} Milgrom, M. 1999, Physics Letters A,
    253, 273
\bibitem []{ospe73} Ostriker, J.P. \&
    Peebles, P.J.E. 1973, ApJ, 186, 467
\bibitem []{pereal99} Perlmutter, S. et al.
    1999, ApJ, 517, 565
\bibitem []{rhs97} Sanders, R.H. 1997, ApJ, 480, 492
\bibitem []{rhs98} Sanders, R.H. 1998, MNRAS, 296, 1009
\bibitem []{rhs99} Sanders, R.H. 1999, ApJ, 512, L23
\bibitem []{rhs00} Sanders, R.H. 2000, MNRAS, 313, 767
\bibitem []{rhs01} Sanders, R.H. 2001, ApJ (in press),
    astro-ph/0011439
\bibitem [] {sv98} Sanders, R.H., Verheijen
    M.A.W. 1998 ApJ, 503, 97
\bibitem []{sz96} Seljak, U. \&
    Zeldarriaga, M. 1996, ApJ, 469, 437
\bibitem [] {sl68} Silk, J. 1968 ApJ, 151, 469
\bibitem []{soleal87} Solomon, P.M., Rivolo, A.R.,
    Barrett, J., Yahil, A. 1987, ApJ, 319, 730
\bibitem []{tf77} Tully, R.B. \& Fisher, J.R.
    1977, A\&A, 54, 661
\bibitem [] {tea96} Tully, R.B., Verheijen, M.A.W.,
    Pierce, M.J., Hang, J-S., Wainscot, R. 1996, AJ, 112, 2471
\bibitem []{vhsc01} Verheijen, M.A.W. \&
    Sancisi 2001, A\&A, 370, 765
\bibitem [] {w72} Weinberg, S., Gravitation and Cosmology (Wiley, New York, 1972)

\end{thebibliography}
\end{document}